\documentclass[runningheads]{llncs}

\pdfoutput=1

\usepackage[T1]{fontenc}

\usepackage[misc,geometry]{ifsym}
\usepackage{color}
\usepackage{graphicx}
\usepackage{listings}
\usepackage{multirow}
\usepackage[ruled,linesnumbered]{algorithm2e}
\usepackage[htt]{hyphenat}
\usepackage[inline]{enumitem}

\begin{document}

%
%
\title{The Nonce-nce of Web Security: an Investigation of CSP Nonces Reuse}

%
%
\author{Matteo Golinelli\inst{1}\textsuperscript{(\Letter)}\orcidID{0000-0002-8743-0825}\and
Francesco Bonomi\inst{1}\and
Bruno Crispo\inst{1}\orcidID{0000-0002-1252-8465}}

\authorrunning{M. Golinelli et al.}

\institute{University of Trento, Trento, Italy\\
\email{matteo.golinelli@unitn.it, francesco.bonomi@hotmail.it, bruno.crispo@unitn.it}}

\maketitle

%
%
\begin{abstract}
Content Security Policy (CSP) is an effective security mechanism that prevents the exploitation of Cross-Site Scripting (XSS) vulnerabilities on websites by specifying the sources from which their web pages can load resources, such as scripts and styles. CSP nonces enable websites to allow the execution of specific inline scripts and styles without relying on a whitelist. In this study, we measure and analyze the use of CSP nonces in the wild, specifically looking for nonce reuse, short nonces, and invalid nonces. We find that, of the 2271 sites that deploy a nonce-based policy, 598 of them reuse the same nonce value in more than one response, potentially enabling attackers to bypass protection offered by the CSP against XSS attacks. We analyze the causes of the nonce reuses to identify whether they are introduced by the server-side code or if the nonces are being cached by web caches. Moreover, we investigate whether nonces are only reused within the same session or for different sessions, as this impacts the effectiveness of CSP in preventing XSS attacks. Finally, we discuss the possibilities for attackers to bypass the CSP and achieve XSS in different nonce reuse scenarios.
\end{abstract}

%
%
\section{Introduction}
\label{sec:introduction}
Content Security Policy (CSP) is a web security mechanism that enables web developers to specify the sources from which their web pages can load resources, such as scripts, style sheets, images, and fonts. CSP is an effective countermeasure to prevent the exploitation of Cross-Site Scripting (XSS) vulnerabilities, which are one of the most common vulnerabilities on the web. CSP is not designed to be the sole mechanism for preventing XSS attacks but it is intended as the last layer of a defense-in-depth approach, providing an additional layer of protection for this type of attack without replacing other mechanisms for preventing XSS vulnerabilities, such as sanitization and validation of untrusted input and the use of templating engines. In practice, while other mechanisms aim at preventing the existence of XSS vulnerabilities, the goal of CSP is to render their exploitation impossible, without eliminating the XSS vulnerabilities.
CSP2 is a W3C specification, published as a Recommendation by the Web Application Security Working Group~\cite{csp2}, and the W3C is currently specifying CSP3 in a Working Draft~\cite{csp3}.
By default, the CSP blocks the execution of all inline scripts and styles (i.e., directly included in the HTML code of a web page). However, CSP2 introduced the concepts of \textit{nonces} and \textit{hashes}, enabling websites to allow the execution of specific individual inline scripts and styles without relying on a whitelist. CSP nonces are a random string included in the policy that is assigned to scripts allowed to execute in the form of an attribute (e.g., \texttt{<script nonce="r4nd0m">}). According to the specification, servers are required to generate a new and unique nonce for each response that includes a policy, and nonces should be at least 128 bits long (before being base64-encoded).

The goal of this study is to measure and analyze the use of CSP nonces in the wild. To achieve this, we performed a large-scale analysis on the Tranco Top 50k sites to detect 1) nonce reuse, 2) short nonces, and 3) invalid nonces.
Reusing the same nonce in more than one response can hinder the protection against XSS attacks, rendering the CSP useless.
We find that more than 10k sites use CSP and, of these, 2271 use CSP nonces. Of the 2271 sites that deploy a nonce-based policy, 598 of them reuse the same nonce value in more than one response.

Next, we analyze the causes of the nonce reuses that we detect to identify whether they are introduced by the server-side code of the website or if the nonces are being cached by a web cache. The strict performance, availability, and scalability standards that websites are required to meet nowadays often lead to the implementation of extremely aggressive web cache configurations, which can induce web caches to erroneously cache dynamic content that they should not, such as CSP nonces. If a web cache caches a page that contains a nonce and serves it to other clients, an attacker can steal it and might use it to bypass the CSP.

Finally, we investigate whether nonces are only reused within the same session (i.e., for a single client) or also for different sessions.
If a website reuses nonces only within a single session, an attacker able to steal its value by exploiting other vulnerabilities is able to bypass the CSP. Instead, if the value of the nonces is the same for all clients, an attacker can simply visit a page of the site to obtain a valid nonce to use to bypass the CSP and perform an XSS attack.

To summarize, we make the following contributions:
\begin{enumerate}
    \item We measure websites' adoption of nonces-based policies on the Tranco Top 50k. We find that 2271 sites use a CSP nonce in at least one of their pages.

    \item We evaluate the implementation of CSP nonces of popular websites in the wild, with a special focus on nonces reuse. To the best of our knowledge, we perform the first large-scale measurement of Content Security Policy nonces reuse, detecting 598 sites that reuse the same value in more than one response.

    \item We investigate the causes of the nonce reuses that we detect, attributing them either to the server-side code of the website or to in-the-middle web caches. Moreover, we analyze if nonce reuses happen only within a single session or also for different ones.
\end{enumerate}

Our code for this research is publicly available as an open-source tool on the author's website~\footnote{https://github.com/Golim/nonce-nce}.

\section{Background}
\label{sec:background}

In this section, we provide an overview of the background knowledge necessary for a thorough understanding of our research.

\subsection{Cross-Site Scripting}
\label{sec:background:xss}
Cross-Site Scripting (XSS) is a type of security vulnerability that enables attackers to inject HTML or JavaScript code into the web pages of a vulnerable website, enabling attackers to execute code in the victim's browsers in the context of the vulnerable pages. Typical injection sources are the query parameters or the path of a request URL. By exploiting XSS vulnerabilities, the attackers can steal victims' information, hijack their sessions and even execute actions on their behalf on the target website.
To prevent XSS vulnerabilities, a common approach is to implement input validation and sanitization to ensure that any user-provided data does not contain any malicious scripts and is properly encoded. Specifically, sanitization removes or HTML-encodes potentially dangerous HTML code from user-provided input to prevent XSS attacks, ensuring that the unsafe content is treated as data and not as code~\cite{xss-prevention}.

\subsection{Content Security Policy}
\label{sec:background:csp}
XSS vulnerabilities are possible because a web browser has no built-in mechanism to determine whether the injected code included in the web pages is reliable or malicious.
The Content Security Policy (CSP) is a security mechanism that helps to prevent XSS attacks, clickjacking, and other code injection attacks by enabling websites to specify what resources can be loaded by the user agent. This mechanism is intended as a defence-in-depth that provides an additional layer of protection against XSS vulnerabilities, by making them not exploitable. CSP can also be used to detect and report code injection attempts, allowing website administrators to take action to prevent further attacks. The CSP can be deployed in \textit{enforcement mode}, meaning that all the violations of the policy will be blocked (and possibly reported) by the browser, or in \textit{report-only mode}. When a policy is report-only, it is not enforced by browsers, which will instead only report CSP violations to a \textit{report-URI} specified in the policy.
The second version of CSP is specified by the W3C in~\cite{csp2}, and they are currently working on its third version (CSP3), published as a Working Draft~\cite{csp3}.
Websites can specify their policy preferably using the \texttt{Content-Security-Policy} response HTTP header, but they can also do it using the HTML \texttt{meta} tag. To deploy a report-only CSP, servers must use the \texttt{Content}-\texttt{Security}-\texttt{Policy}-\texttt{Report}-\texttt{Only} response header, and cannot do it using the meta tag.
The CSP works by allowing website administrators to specify a whitelist of trusted sources for executable content, such as JavaScript, CSS, and images. Any content that comes from untrusted sources must be blocked by the user agent~\cite{csp2}.
A CSP is composed of one or more directives separated by a semicolon, and each directive is used to specify the valid sources for a particular type of resource. These directives can be used to whitelist specific domains or sources from which resources can be loaded, and to block the ones that do not match the specified sources. The \texttt{default-src} directive is used as a fallback, specifying the default behaviour for any directive that is not explicitly specified. By default, the CSP blocks all inline scripts and styles, effectively preventing the exploitation of XSS vulnerabilities.

\subsubsection{Nonces}
\label{sec:background:csp:nonces}
The second version of this security mechanism, CSP2, introduced the concept of \textit{nonces} and \textit{hashes} as a way of allowing the execution of individual inline scripts and styles without relying on a whitelist.
The concept of ``nonces'' was initially proposed by Needham et al. in 1978 and specifically means a number ``used only once''~\cite{needham1978}.
A \textit{CSP nonce} is a random value used only once, which is added to the inline scripts or style tags of a web page as an attribute. The nonce is included in the policy, which tells the browser to only execute scripts or styles that have a matching nonce value~\cite{csp2}. For example, the following policy:

\begin{lstlisting}
Content-Security-Policy: default-src 'self';
    script-src 'nonce-cmFuZG9t' 'self';
\end{lstlisting}

will allow the execution of the first inline script in the following example, but will block the second and the third scripts because the nonce is missing or invalid, respectively.

\begin{lstlisting}[language=HTML]
<script nonce="cmFuZG9t">
    console.log("This will execute");
</script>
<script>
    console.log("This will *not* execute");
</script>
\end{lstlisting}

The CSP2 specification~\cite{csp2} indicates that a nonce should have the following characteristics:
\begin{enumerate*}
    \item Must be unique for each HTTP response that includes a CSP.
    \item Should be generated using a cryptographically secure random number generator.
    \item Should be at least 128 bits long, before encoding.
\end{enumerate*}

A CSP nonce should be long and randomly generated to prevent attackers from guessing or brute-forcing it.
However, it must be noted that properly implementing a nonce policy does not necessarily prevent the exploitation of XSS vulnerabilities altogether. In fact, if an inline script with the correct nonce attribute uses untrusted user input, an attacker could bypass the policy and achieve XSS.
If a CSP nonce is reused for multiple HTTP responses, an attacker who is able to steal it could effectively bypass the policy by injecting a script with the correct nonce value.

\subsection{Web Caches}
\label{sec:background:web_caches}
Web caches are an essential component of modern web architecture, enabling faster and more efficient web browsing by storing previously accessed resources. Moreover, caches are generally physically located closer to the user's location, reducing the latency. When a user requests a web page or resource that is already in the cache, the cache can deliver it much faster than if it had to be fetched from the origin server. Caching also reduces the load on origin web servers. Web caches can be implemented at various stages of the web architecture, including the client side, proxy servers, and Content Delivery Networks (CDNs).

\paragraph{Web Cache Deception}
Web Cache Deception (WCD) is a vulnerability that enables attackers to induce a public cache into caching information that it should not store, resulting in that information being publicly available.
Mirheidari et al. in~\cite{wcde} showed how WCD vulnerabilities can be exploited to force a web cache into mistakenly caching a CSP nonce. If the cached nonces are reused in the same session for multiple requests coming from the origin server, an attacker could bypass the CSP and perform an XSS attack as follows:
\begin{enumerate}
    \item The attacker exploits a WCD vulnerability to steal a valid CSP nonce linked to the victim's session.
    \item The attacker crafts an XSS payload including the stolen CSP nonce in the injected script.
    \item The attacker induces the victim into visiting a web page or following a malicious URL.
\end{enumerate}

\paragraph{Cache status headers}
Web caches generally employ a response header to communicate whether a resource that they handle is coming from the origin web server or is a cached copy. Different web cache technologies might use different header names and values for this purpose, but the majority of them use the keyword \texttt{cache} in the header name, and the keywords \texttt{HIT} and \texttt{MISS} to indicate a cached resource and one coming from the origin, respectively.

\section{Related work}
\label{sec:related_work}

The Content Security Policy was proposed by Stamm et al. in 2010, as an additional layer of security for XSS and CSRF attacks~\cite{stamm2010}.
Numerous studies have measured the adoption of CSP by websites in the wild and its variation over time, giving us an indication of how much its use has grown through the years.
In 2014, Weissbacher et al. measure the adoption of CSP in the Alexa Top 1M over a period of 16 months and find that only 850 of them use it~\cite{weissbacher2014csp}. Calzavara et al. do the same in 2016~\cite{calzavara2016content} and 2018~\cite{calzavara2018}, finding respectively 8,133 and 16,353 sites using the CSP, marking a significant increase in its adoption.
Finally, Roth et al. show that, between 2012 and 2018, 1,233 of the 10,000 websites analysed deployed a CSP for at least one day~\cite{roth2020complex}. Roth et al. also measure for the first time the adoption of CSP nonces and hashes, finding that in 2018 only 5\% of sites adopting a CSP used nonces and 1\% hashes~\cite{roth2020complex}.

Over the years, numerous studies have been presented presenting techniques to automate CSP policy generation.
In 2013, Doup\'{e} et al. present deDacota, an automatic tool to statically rewrite code to separate data from the code and enforce this separation at run-time using CSP~\cite{deDacota2013}.
The following year, Johns analyses the dangers of dynamically filling scripts with data retrieved at run-time and proposes a templating and checksumming mechanism that enables servers to communicate which scripts are allowed to run to the browser~\cite{johns2014}. 
Weissbacher et al. compare the adoption of CSP with other security headers and experiment with a semi-automated crawler-based CSP generation mechanism~\cite{weissbacher2014csp}.
In 2016, Kailas and Braun analyse the usage of CSP in real-world websites, identifying errors and inconsistencies, and present a tool to automatically generate policies on the client side~\cite{kailas2016}.
Pan et al. propose \textit{CSPAutoGen}, a tool that auto-generates policies in real-time according to the page content and the templates of scripts, without requiring modifications to the server-side code~\cite{pan2016}.
Kerschbaumer et al. show that 90\% of CSP deployments include the \texttt{unsafe-inline} keyword, rendering it ineffective in preventing XSS attacks, and propose a system to automatically generate CSPs by whitelisting only the hashes of expected scripts~\cite{Kerschbaumer2016InjectingCF}.

In 2015, Hausknecht et al. analyse the interplay between the CSP and browser extensions, finding that some extensions tamper with the CSP of websites to be able to work~\cite{hausknecht2015may}.
In 2017, Some et al. analyse how the SOP could cause CSP violations~\cite{some2017}, and in 2021 Steffens et al. analyse how the use of third-party resources impacts a website's implementation of CSP and argue that relying on third parties is a major roadblock for security~\cite{steffens2021WhosHT}.
In 2021, Roth et al. present a qualitative study that analyses the difficulty of developing a safe CSP by tasking real-world developers with a programming task, where they have to develop a CSP for a small web app to prevent XSS attacks~\cite{roth2021}.
Calzavara et al. in 2016 show that CSP has limited deployment and that the deployed policies exhibit several weaknesses and misconfiguration errors~\cite{calzavara2016content}.
In 2017, Calzavara et al. propose an extension to CSP called \textit{Compositional CSP} that enables the composition of policies at run-time and assess its potential impact in the wild~\cite{calzavara2017ccsp}.
In 2016, Van Acker et al. show how to exploit DNS prefetching to exfiltrate data, bypassing the protection offered by the CSP~\cite{vanacker2016}.
In 2016, Weichselbaum et al. show that only 0.16\% of the domains they analysed use CSP and detect that 94.7\% of distinct policies deployed can be bypassed. They argue that maintaining a secure whitelist for a complex application is infeasible in practice, and suggest replacing URL whitelisting with nonces and hashes~\cite{weichselbaum2016csp}.
In 2017, Lekies et al. present a novel attack that exploits \textit{script gadgets}, i.e., small fragments of JavaScript in a site's legitimate code that can be used to bypass XSS protection mechanisms, including the CSP. They identify gadgets in at least 19.9\% of the tested sites~\cite{lekies2017}.

Finally, we acknowledge the valuable contributions made by a separate team of researchers who explored a closely related problem in parallel with our research. Trampert et al. present an investigation of CSP nonces reuse, underscoring the significance of this topic~\cite{trampert2023honey}. To attribute nonces reuse to a cache, they check whether a website uses a CDN; however, this approach is susceptible to generating false positives, given that the mere usage of a CDN by a website does not necessarily guarantee that the present response is being cached. In our work, we use cache busting and lookup of the cache status headers to more accurately attribute nonce reuses to web caches.

\section{Methodology and Experiment}
\label{sec:methodology}
Our methodology is composed of three main phases: 1) URLs Collection, 2) CSP Nonces Detection, and 3) CSP Nonces Evaluation.

\subsection{URLs Collection}
\label{sec:methodology:crawling}
In the first phase, we crawl the domain name of a website in an unauthenticated way to identify the URLs that present a CSP. Specifically, we visit the homepage of the website and then recursively visit all the links that we find in the source code of the pages. We only follow internal URLs, i.e., URLs where the domain is either the same as the one on the website or a subdomain. Our crawler can be configured to only visit a maximum number of subdomains and a maximum number of pages for each subdomain. 
Our crawler is developed in Python and uses the \textit{BeautifulSoup} library to extract the links from the HTML code of web pages, and the \textit{requests} library to perform HTTP requests.

\paragraph{CSP Detection}
To identify the URLs on a website that present a CSP, we look for both the \texttt{Content-Security-Policy} header and the \texttt{meta} tag in the HTML of the page. We also test whether the CSP is deployed in report-only mode by checking the presence of the \texttt{Content-Security-Policy-Report-Only} header (a report-only CSP cannot be deployed using the \texttt{meta} tag).

\subsection{CSP Nonces Detection}
\label{sec:methodology:nonces_detection}
In this phase, we visit all the URLs that present a CSP collected in the previous step and check if they use CSP nonces as part of their policy. Since we are interested in sites where nonces are actually used, we do not only check if they include a nonce in their policy, but we verify if the page includes \texttt{script} tags with the \texttt{nonce} attribute. We save the HTML source code of the pages that include a CSP nonce and we store the headers of the requests and the responses for future analyses.

\subsection{CSP Nonces Evaluation}
\label{sec:methodology:nonces_evaluation}
In this phase, we check if the length of the previously identified CSP nonces is sufficient and if they are reused.
To detect nonce reuse, we request the pages that include a nonce a second time and check if the value of the nonce is the same. If the nonce value is the same in both responses, we mark the page as vulnerable to nonce reuse.
Then, we mark a nonce as too short if it is shorter than 22 useful characters (i.e., not including ``\texttt{=}'' padding characters) in its base64-encoded form.

\subsubsection{Reuse Causes}
When we detect nonce reuse, we analyze whether it is introduced by a web cache caching a dynamic nonce, or if the reuse is caused by the server-side code of the website. To do this, we employ three mechanisms that, used together, give us a strong indication of whether a cache is influencing the nonces or not.

\paragraph{Static Nonces}
First, we attribute to the server-side code all the nonce reuses where the nonce is the same in all responses coming from a website (only on the sites that use a nonce on two or more pages).
It must be noted that these nonces might still get cached, but the root cause of their reuse is the server-side code.
Next, we use two techniques explained below called \textit{Cache-Busting} (CB), and \textit{Cache Header Heuristics} (CHH). We perform this test because, if a nonce is reused due to a web cache storing it, an attacker is not able to perform an XSS attack even if they could steal the nonce, as the cached response does not include the attacker's injection. Assuming that the nonce is reused only due to a cache, the attacker is not able to overwrite the cached copy of the response without the nonce changing.

\paragraph{Cache-Busting}
For this test, we perform a third request including a cache-busting query parameter in the URL (i.e., a randomly generated parameter added to the query string of the URL). This technique is effective when the query string is included in the cache key (i.e., the unique identifier of a resource stored by a cache~\cite{cache_key}). By adding a random modification to the query string, we cause our request to have a different cache key, effectively preventing a cache from serving us a cached response. If the response to the cache-busted request includes a different nonce, we can attribute the nonce reuse to a web cache. However, if the value of the nonce is the same as the one in the previous responses, we cannot conclusively exclude the presence of a cache, as some web caches might not include the query string in their cache key, resulting in our cache-busting mechanism failing in its purpose of excluding the cache. For this reason, we also employ the \textit{CHH}.

\paragraph{Cache Header Heuristics}
In this test, we check whether the response with a reused nonce is coming from a cache or the origin server using the \textit{cache header heuristics} algorithm presented by Mirheidari et al. in~\cite{wcde}~\footnote{The code of the algorithm can be found at \url{https://github.com/golim/wcde}}.
This test performs a lookup of the cache status headers of the second response (i.e., where we detected nonce reuse) to check if it is coming from a cache or the origin server, as described in Section~\ref{sec:background:web_caches}. If we detect that the second response is not coming from a cache, we can conclusively exclude the influence of a cache on the nonce reuse. Otherwise, if the second response is coming from a cache or if the test gives an unknown status (e.g., because the cache status headers are missing in the response), we cannot conclusively attribute the nonce reuse to a cache or not, and we need to resort on the \textit{cache-busting} test. If we detect that the second response is coming from a cache, we cannot exclude that the server-side code is nevertheless issuing the same nonce.

To summarize, only by using all three tests in combination we can attribute nonce reuse to a cache or not with a high degree of confidence.

\subsubsection{Session Analysis}
Finally, to test whether nonces are only reused for the same session or also for different ones, we perform another HTTP request, without providing the previously stored cookies.

\subsection{Experiment}
\label{sec:methodology:experiment}
We performed a large-scale experiment on the web to detect CSP nonces reuse and websites that use short nonces. Our dataset is composed of the Tranco Top 50k~\cite{tranco} downloaded on 21, July 2022~\footnote{The specific dataset that we used can be downloaded at \url{https://tranco-list.eu/list/W97W9/}}. For each domain in our dataset, we crawled at most 10 pages in at most 10 subdomains (i.e., a maximum of 100 pages for each domain in our dataset). For each domain that we tested, we created a session object with cookies persistence, simulating a user surfing a website using the same browser for all requests.

\section{Results}
\label{sec:results}
In this section, we discuss the results of our large-scale measurement over the Tranco Top 50k.

\begin{table}
    \caption{The number of sites that deploy a Content Security Policy, use CSP nonces, and reuse the same nonce value for multiple responses. Percentages are calculated over the total number of sites that deploy a CSP (10034).}
    \label{tab:results}
    \centering
    \setlength\tabcolsep{3pt}
    \begin{tabular}{lrrr}
    \hline

    \textbf{Total sites using CSP} &
    \textbf{10034} & {}
    \\

    \hline

    \multicolumn{1}{r}{\textit{enforcement mode}} &
    8946 & (89.2\%)
    \\

    \multicolumn{1}{r}{\textit{report-only mode}} &
    1088 & (10.8\%)
    \\

    \hline

    Sites with CSP nonces &
    2271 & (22.6\%)
    \\

    Sites reusing CSP nonces &
    598 & (6.0\%)
    \\

    \hline
    \end{tabular}
\end{table}

\subsection{CSP Adoption and Usage}
Table~\ref{tab:results} summarizes the results of our measurement of CSP adoption.
Even though it is not the main focus of our research, we briefly analyze our measurement of the adoption of CSP. We crawled 50k websites and detected 10034 (20.1\%) deploying a Content Security Policy. Of these, 1088 sites only use the CSP in report-only mode, while 8946 enforce their policy. We also found 346 sites that only deploy their policy in the \texttt{meta} tag, while 135 deploy it both in the \textit{Content-Security-Policy} header and the \texttt{meta} tag.

\subsection{Reuse Analysis}
\begin{figure}[t]
  \centering
  \includegraphics[width=0.7\linewidth]{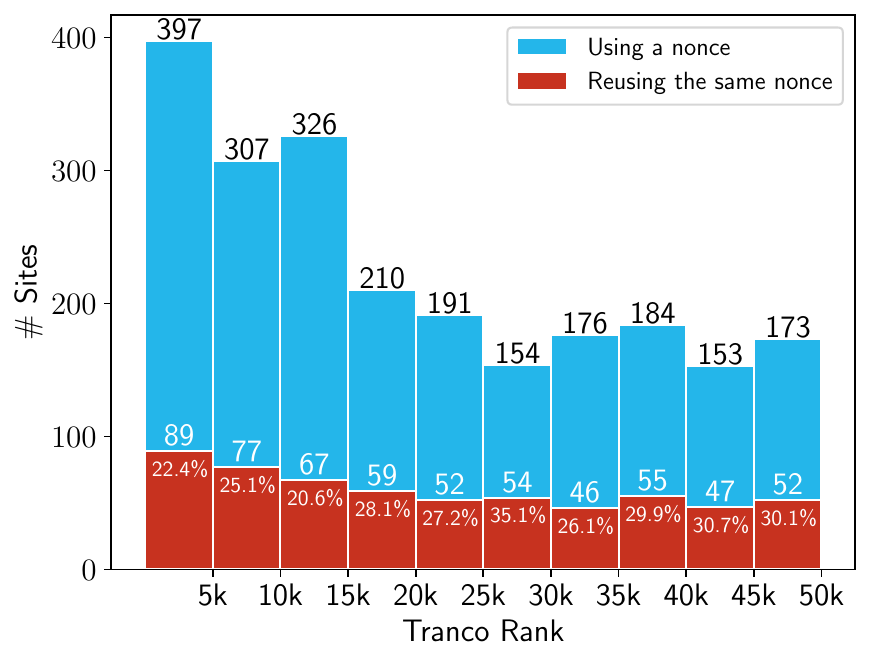}
  \caption{The distribution of websites that have a nonce-based CSP (in blue),
    and the subset of those which reuse a CSP nonce (in red)
    with respect to their ranking in the Tranco Top 50k.
    The percentage in each bar is calculated over the number of sites that use a nonce in the same 5k bucket.}
  \label{fig:distribution}
\end{figure}

Table~\ref{tab:reuse} presents the results of our measurement of websites that use a CSP and that reuse nonces. Of the 10034 identified sites that deploy a CSP, 2271 (22.6\%) present a nonce-based CSP in at least one of the pages that we visited. 598 (26.3\% of the sites using a nonce-based CSP) reused the same CSP nonce value in multiple responses. Fig.~\ref{fig:distribution} shows the distribution of websites that use a nonce CSP in at least one of their pages and the sites that reuse at least one nonce, with respect to their ranking in the Tranco Top 50k. From Fig.~\ref{fig:distribution}, we can see that the usage of CSP nonces is higher among the more popular websites according to their ranking, while the percentage of websites that reuse the same nonce value in more than one response is higher among the lower rankings.

\begin{table}[t]
    \caption{The number of sites presenting at least one reused CSP nonce and the investigated reason for the reuse.
    Percentages are calculated over the total number of sites that reuse a nonce (\textit{598}).}
    \label{tab:reuse}
    \centering
    \setlength\tabcolsep{3pt}
    \begin{tabular}{lrrrrrr}
    \hline
    
    \textbf{Total sites reusing nonces} &
    \textbf{598} & {}
    \\

    \hline

    \multicolumn{1}{r}{\textit{due to a cache}} &
    256 & (42.8\%)
    \\

    \multicolumn{1}{r}{\textit{server-side code}} &
    342 & (57.2\%)
    \\

    \hline

    \multicolumn{1}{r}{\textit{in the same session}} &
    37 & (6.2\%)
    \\

    \multicolumn{1}{r}{\textit{in different sessions}} &
    561 & (93.8\%)
    \\

    \hline
    \end{tabular}
\end{table}

\subsubsection{Causes Analysis}
We investigated the effect of web caches on the reuse of the nonce and found that 342 (57.2\% of the 598 sites reusing a nonce) sites use the same value for nonces in more than one origin server response, indicating that the cause of reuse is code executing on the site's server, while for 256 (42.8\% of the 598 sites reusing a nonce) sites the cause of nonce reuse is solely due to a cache storing a copy of the response that includes an otherwise dynamic nonce and serving it in response to the subsequent requests. 219 sites use the same nonce value in all responses coming from the origin server.

\paragraph{Cached Nonces}
Using cache-busting and the cache header heuristics, we also detected 318 websites that cache CSP nonces (53.2\% of the sites that reuse nonces), regardless of the nonces being static or dynamic when generated by the server-side code. Of these 318 cases where we observed caching, 190 were detected using cache-busting, and 128 using the CHH algorithm.

\subsubsection{Sessions Analysis}
Our goal was to detect websites that reuse identical nonce values within a session and those that reuse the same nonce across all visitors to the site. As described in Section~\ref{sec:methodology}, to do this we performed the same request using a clean browser, to simulate a different user visiting the same page. We detected that only in 37 cases the nonce value was bound to the session of the visitor, while in the vast majority of the cases (561), we observed that the same nonce value was repeated for visitors with a different session (i.e., with different cookies). It is important to note that, when a reused nonce value is not bound to the session of the visitors, an attacker could simply visit the same page or site to obtain a valid CSP nonce that will be used in the responses to the victim requests, effectively rendering the CSP useless to protect against XSS attacks.

We performed all the tests using the same IP address and, even if we think this is highly unlikely, there is the possibility that some sites bind the nonce value to an IP address. However, even if this is the case, an attacker in the same Local Area Network (LAN) as the target victim would still be able to obtain a valid CSP nonce value to perform an XSS attack.

\subsection{Length and Validity Analysis}
We analyzed all the nonces detected in the web pages that we visited to identify the ones shorter than the specification recommends. It is important to note that it is extremely difficult for an attacker to perform a brute-force attack against shorter CSP nonces, but we still think that this analysis can be interesting to highlight how many websites do not follow the W3C specifications.
Of the 2271 websites using a nonce, 501 (22.1\%) use a CSP nonce shorter than recommended. Interestingly, 356 (15.7\%) use a nonce of length 8.
Moreover, we identified 8 websites that include not valid characters (i.e., characters not included in the \textit{base64url-encoding} alphabet). All modern browsers reject these nonces and block the execution of inline scripts.

To summarize, we first measured the sites deploying a CSP, using nonces, and reusing the same value more than once in Table~\ref{tab:results}. Next, in Table~\ref{tab:reuse}, we analyzed the possible causes for nonce reuses and whether they happen only within a single session or also for different ones. Finally, we analyzed the length and validity of the nonces.

\section{Discussion}
\label{sec:discussion}
In this research, we measured several aspects of the adoption of the Content Security Policy by popular websites. We found that, compared to previous large-scale measurements, the adoption of CSP and nonces-based policies has grown. However, our research highlights that more than one in every four sites that use nonces misuses them, repeating the same value in more than one response (Table~\ref{tab:results}). Reusing nonces could result in a complete bypass of the CSP, allowing attackers to exploit possible XSS vulnerabilities.
We analyzed the different conditions in which nonces reuse happens and investigated its possible causes in Table~\ref{tab:reuse}. Depending on these, it is easier or harder for an attacker to bypass CSP to exploit XSS vulnerabilities.

Specifically, if the same nonce value is reused for different sessions, it is effectively the same as deploying an \texttt{unsafe-inline} directive (i.e., allowing the execution of all inline scripts and styles, regardless of their origin). In fact, an attacker can simply visit the vulnerable website to obtain a valid CSP nonce value to craft an injection that bypasses the CSP.
On the other hand, if a nonce is only reused within the same session, the attacker has to exploit other vulnerabilities or attacks to steal the nonce value from the target victim. For example, an attacker can exploit possible Web Cache Deception vulnerabilities or perform cross-origin requests (if the \textit{SameSite} attribute of the cookies is set to \textit{None}). Once the attacker holds a nonce linked to the session of a victim, they can craft an injection that includes the stolen nonce to exploit an XSS vulnerability.
Next, we analyzed the causes of nonce reuses to detect the ones likely caused by the presence of a web cache storing otherwise dynamic nonces. As described in Section~\ref{sec:methodology:nonces_evaluation}, to do this we used three approaches: cache-busting, cache header heuristics (i.e., look-ups of the response headers), and the comparison of nonces on different pages of the same site. If a nonce is reused due to a cache, it is harder for an attacker to bypass the CSP and achieve XSS.
In fact, even if an attacker can steal a cached nonce of a victim, they cannot inject a malicious payload in the cached copy of the response directly.
An attacker can achieve XSS under these conditions if they can inject a payload with the stolen nonce exploiting client-side XSS vulnerabilities, i.e., where untrusted data is dynamically loaded directly from the victim's browser (e.g., due to DOM XSS, or when the page includes data dynamically through XMLHttpRequest or AJAX).
We analyzed the length of the nonces and found that more than a fifth of all sites that employ nonces have them shorter than recommended by the specification. Depending on the length of the nonces, brute-force attacks might be feasible or not.
Finally, we tested the validity of the nonces by checking that they only contained valid characters, finding that 8 sites presented invalid nonces. Browsers reject such nonces and block the execution of all the scripts and styles, resulting in a self-Denial of Service.

\subsection{Reuse Causes}
Here we discuss the possible causes for nonce reuse (see Table~\ref{tab:reuse}). Each web request response is generally handled by several entities with different functions (e.g., caches, proxies, firewalls, origin servers). The use of \textit{clean URL} techniques (i.e., URLs whose structure does not directly reflect the file system structure of the server) means that often only the origin server understands the true nature of a resource, while the abstraction introduced by these techniques makes it obscure to other in-the-middle entities. It is therefore possible for a cache to be configured in such a way as to mistakenly cache pages that include dynamic content, such as CSP nonces.
When instead a nonce reuse is not caused by a cache but by the server-side code of the website, it is more difficult to attribute the cause to a specific entity. Indeed, this may be caused by numerous factors that cannot be investigated from the outside in a black-box manner.
Previous studies have focused on the complexity of implementing a proper CSP that effectively protects against XSS attacks~\cite{roth2020complex}, even interviewing web developers directly~\cite{roth2021}. Some possible causes of these nonce reuses may be a lack of knowledge of the technology and possible attacks of developers, programming errors, typos, and negligence.

\subsection{Limitations and Future Work}
In this section, we acknowledge the limitations of our research and pave the way for future works to further explore this area. First, we do not investigate the randomness of the nonces. A possible way to do it would be to collect a certain number of nonces from each site and perform an analysis of their entropy.
In our work, we also do not perform any analysis on the scripts included in the pages to check if they use untrusted data, hindering the security of the CSP in preventing XSS attacks even if the nonces are used correctly. Additionally, we do not test for the presence of XSS vulnerabilities because that is out of the scope of this paper. However, we collected all the HTML code of the pages that include a nonce, and we will perform this analysis in future work.
Finally, even by using both the cache header heuristics and the cache-busting tests, we cannot attribute nonce reuse to a cache or the server-side code with 100\% certainty. In fact, detecting the presence of a cache is a complicated task that, to the best of our knowledge, cannot be performed with full certainty of the results. According to the state-of-the-art, using both these tests is the way to most accurately interpret the data that we collected.

\section{Conclusion}
Our study highlights the importance of correctly implementing Content Security Policy (CSP) nonces to prevent Cross-Site Scripting (XSS) attacks. Our analysis of the Tranco Top 50k sites revealed that many websites reuse the same nonce value multiple times, which can potentially allow attackers to bypass CSP protections.
We analyzed the different scenarios of nonce reuse and discussed the possibilities for the attackers to successfully bypass the CSP. Specifically, we investigated the root causes of nonce reuses, attributing them either to the server-side code, or in-the-middle web caches. When nonce reuse is only due to a cache, it is harder for an attacker to bypass the CSP, as they have to find a way to inject the malicious payload in the cached web page dynamically, for example exploiting DOM XSS vulnerabilities.
If instead, the server-side code issues the same nonce for multiple responses, an attacker can simply use a stolen nonce to craft a payload that includes it to bypass the CSP. In this scenario, the complexity of the attack depends on whether the nonce is reused only within a single session or also for different ones. In the first case, the attacker has to steal a nonce valid for the session of the victim exploiting other vulnerabilities. We showed how, for example, this can be done by exploiting Web Cache Deception vulnerabilities or, in specific conditions, making cross-site requests. If instead the same value is reused as a nonce for arbitrary sessions on a website, the attacker can obtain a valid nonce simply by visiting the website. This type of nonce reuse is practically the same as enabling all inline scripts with the \texttt{unsafe-inline} directive.
To conclude, reusing the same nonce value is a dangerous behaviour by websites that can effectively hinder the protection offered by the CSP from the exploitation of XSS vulnerabilities.
Reusing the same nonce is in some cases the same as allowing all scripts inline, in others, it is a severe relaxation of policy with a dramatic reduction in the protection offered.
Implementing a proper nonce-based policy is a complex and costly task, but it is the only way a website using it can fully protect itself against XSS.

%
%
\subsubsection{Acknowledgements}
This work has been partially supported by the EU Horizon project DUCA (GA 101086308) and CrossCon (GA 101070537). Views and opinions expressed are however those of the author(s) only and do not necessarily reflect those of the European Union or CINEA. Neither the European Union nor the granting authority can be held responsible for them.

%
%
\bibliographystyle{splncs04}
\bibliography{paper}

\end{document}